\documentstyle[aps,prl,preprint,tighten,epsfig]{revtex}
\begin{document}
\title{Kondo effect in a one dimensional d-wave superconductor}
\author{Karyn Le Hur}
\address
{Theoretische Physik, ETH-H\"onggerberg, CH-8093 Z\"urich, Switzerland}
\date{\today}
\maketitle

\begin{abstract}
We derive a solvable resonant-level type model, to describe 
an impurity spin coupled to zero-energy bound states localized
at the edge of
a one dimensional d-wave superconductor. This results in a
two-channel Kondo effect with a quite unusual low-temperature thermodynamics.
For instance, the {\it local} impurity susceptibility yields a finite
maximum at zero temperature (but no $\ln$-divergence) due to the splitting
of the impurity in two Majorana fermions. Moreover, we
make comparisons with the Kondo effect
occurring in a two dimensional d-wave superconductor.
bipolarons, resonating valence bond model, anyon mechanism, marginal Fermi
liquid, Luttinger liquid, etc.).
\end{abstract}

\pagebreak

\narrowtext

{\it Introduction and motivation}.--- 
The problem of magnetic impurities in a d-wave superconductor has been
extensively discussed in the literature. This problem is of relevance for
high-$T_c$ and heavy-fermion superconductors.
Since the pioneering paper by Abrikosov and Gor'kov\cite{Abrikosov} it is known
that magnetic impurities act as pair-breaking centers
\cite{Ueda_1985,T_c_experim} and
hence reduce the amplitude of the condensate.
Recently, it
has been shown that quantum mechanical fluctuations of magnetic impurities
give rise to other important effects, especially when coupled to the fermionic
quasiparticles of the superconducting state. In a conventional metal, these
correlations lead to the so-called Kondo effect\cite{Hewson}, i.e. 
the complete screening
of the impurity at low temperature. In d-wave superconductors, the density
of normal quasiparticles vanishes at the Fermi level. Then, screening is
absent in perturbation theory and a {\it critical} exchange coupling between
the quasiparticles and the magnetic impurity is crucial for Kondo screening
to take place\cite{Cassanello_d-wave_Kondo}. 
Using an effective model of one-dimensional chiral
fermions coupled to a single impurity and 
the
so-called large N-expansion (N, being the rank of the symmetry group of the
impurity spin), Cassanello and Fradkin have shown that
there is a {\it quantum phase transition}\cite{QP} from an 
unscreened
impurity state to an overscreened Kondo regime\cite{Nozieres_intro}.
This occurs at a critical 
value $J_c$ of the coupling constant
which is typically a fraction of
$\Delta$, the size of the gap away from the nodes at zero temperature. In 
the overscreened phase, which 
arises due to the consequent coupling between normal
quasiparticles close to one particular node
on the Fermi surface and the impurity, the low-temperature
thermodynamics is given by $\chi_{imp}\sim T\ln T$ and
$C_{imp}\sim T^2\ln T$. In this regime 
the Kondo scale $T_k$ is almost always smaller
than $\Delta$, 
such that pair-breaking effects should not be prominent\cite{Borkowski}.

On the other hand, the spin liquid phase of the two-leg ladder system
gives us a simple model of a Mott-insulating phase which actually exhibits
pairing, with an approximate d-wave symmetry. Moreover, upon doping, the
two-leg ladder yields quasi-long-range superconducting (d-wave) pairing
correlations\cite{Schulz,Fisher_les Houches}. 
Experimentally, superconductivity has up to now only been observed
in a two-leg ladder material under high pressure \cite{two-leg-exp}.
In this Letter, we investigate for the first time the (two-channel) Kondo
model occurring in a two-leg ladder, for a {\it weak} 
repulsive electron-electron
interaction $U$. 
First, based on bosonization methods and previous
results,
we derive an effective 
Hamiltonian (called below, two-band Luther-Emery Hamiltonian)
describing well a two-leg ladder d-wave superconductor at zero
temperature. Second, we show that the Emery-Kivelson solution of the usual
two-channel Kondo model\cite{EK} and the effective 
two-band Luther-Emery Hamiltonian, can be combined leading
order in $U$. The model becomes
solvable and the low-temperature thermodynamics can be computed
explicitly. Additionally, an exact solvable model for an impurity coupled to
a one-dimensional s-wave superconductor has been proposed in
ref.\cite{Gogolin}. In
comparison, the problem of a magnetic impurity in Luttinger
liquids has been more studied\cite{Furusaki,KLH}.

{\it Model}.--- To study low-energy physics, we use a continuum limit 
formalism ($a$ is a required short-distance cutoff). 
The noninteracting two-leg ladder of spinful fermions
is described by the Hamiltonian density
\begin{equation}
{\cal H}_o=-t\sum_{j;\alpha}
\Psi^{\dag}_{j\alpha}(x+a)\Psi_{j\alpha}(x)
+h.c.-t_{\perp}\sum_{\alpha}\Psi^{\dag}_{1\alpha}(x)\Psi_{-1\alpha}(x)
+h.c.,
\end{equation}
where t ant $t_{\perp}$ are the hopping amplitudes along and between the
chains and $\Psi_{j\alpha}(x)$ creates a fermion with spin 
$\alpha=\uparrow,\downarrow$
in the chain j at the rung x. In what follows, we are going
to consider the limit $U\ll t$ and $t_{\perp}\sim t$. Therefore, we
can diagonalize first ${\cal H}_o$
in the so-called bonding-antibonding band basis, defining: 
$\Psi_{o\alpha,\pi\alpha}=
1/\sqrt{2}(\Psi_{1\alpha}\pm\Psi_{-1\alpha})$. 
The bonding $\Psi_{o\alpha}$ and the antibonding
$\Psi_{\pi\alpha}$ operators are expanded as $(j=o,\pi)$:
$\Psi_{j\alpha}\sim \Psi_{Rj\alpha}e^{ik_{Fj}x}+\Psi_{Lj\alpha}e^{-ik_{Fj}x}$.
Upon linearizing the spectrum around the
four Fermi points the kinetic energy takes the form,
\begin{equation}
{\cal
  H}_o=-v\sum_{j=o,\pi;\alpha}[\Psi^{\dag}_{Rj\alpha}i\partial_x
\Psi_{Rj\alpha}-\Psi^{\dag}_{Lj\alpha}i\partial_x\Psi_{Lj\alpha}].
\end{equation}
Moreover, not so far from half-filling
the Fermi velocity in each band is the same, hereafter denoted as 
$v$\cite{Hubl}. Now, 
we must include the 
effects of the electron-electron interactions. As a generalization of
the Hubbard interaction, the two-leg ladder is described by 
a set of marginal Fermion interactions 
(consult Eq.(2.5) of ref.\cite{Fisher_les Houches}),
that can be readily decomposed in terms
of the continuum Dirac fields. 
Remarkably, the
renormalization group equations reveal that {\it before} the flows of 
the different parameters
scale towards strong coupling, they are attracted to a special direction,
characterized by a unique coupling constant $g\sim U$. Low-energy physics
depends only on this coupling. To describe it precisely we need to
``bosonize'' the electron fields as follows\cite{Haldane}
\begin{equation}
\label{field}
\Psi_{Pj\alpha}=\kappa_{j\alpha}e^{i(P\phi_{j\alpha}+\Theta_{j\alpha})},
\end{equation}
with $P=R/L=\pm$ and $\kappa_{j\alpha}$ the corresponding Klein factors.
The displacement field $\phi_{j\alpha}$ and phase field
$\Theta_{j\alpha}$ satisfy the usual commutation rules. It is convenient to
separate charge and spin modes, defining: $\phi_{jc}=(\phi_{j\uparrow}+
\phi_{j\downarrow})/\sqrt{2}$ and $\phi_{js}=(\phi_{j\uparrow}-
\phi_{j\downarrow})/\sqrt{2}$. The Hamiltonian density 
reads (see Eqs.(3.2) and (3.4) of Ref.\cite{Fisher_les Houches}): ${\cal 
H}={\cal H}_o+{\cal H}_I$, with
\begin{eqnarray}
\label{int}
{\cal 
H}_o&=&\frac{1}{2}\sum_{a=jc,js}[\frac{v}{K}(\partial_x\phi_{a})^2+vK(\partial_x
\Theta_{a})^2],
\\ \nonumber
{\cal H}_I&=&-g\cos\sqrt{2}(\Theta_{oc}
-\Theta_{\pi c})\cos\sqrt{2}\phi_{os}\cos\sqrt{2}\phi_{\pi s},
\end{eqnarray}
$K=1-g/\pi v<1$, being the Luttinger parameter of the 
effective theory. ${\cal H}_I$ is produced chiefly by {\it interband}
charge and spin Cooper backward scatterings.

When the single coupling
constant $g$ is flowing off to strong couplings, the two spin fields
$\phi_{os}$ and $\phi_{\pi s}$ will be pinned in the minima of cosine
potentials, i.e. $\phi_{js}\simeq 0$. The spin in each band 
vanishes\cite{newr}. 
Furthermore, the coherence between 
the bands arises due to the pinning of  
$\Theta_{cf}=(\Theta_{oc}-\Theta_{\pi c})/\sqrt{2}\simeq 0$
that guarantees the binding between `holons' of the 
two bands\cite{flavor-inter}: the 
superfluid system is ruled by the unique phase
$\Theta_{c}=(\Theta_{oc}+\Theta_{\pi c})/\sqrt{2}=\sqrt{2}\Theta_{jc}$.
As a consequence, the
pair field operator $\Delta_j=\Psi_{Rj\uparrow}\Psi_{Lj\downarrow}
\sim \pm\exp i\Theta_{c}$ yields only a very slowly 
decreasing power-law decay: $<\Delta_j^{\dag}(x)\Delta_j(0)>\ \sim x^{-1/2K}$. 
This produces prominent
d-wave superconductivity\cite{dlike}. Since $K<1$, the 
(marginal) operator in ${\cal H}_I$ is relevant and then
generates a superconducting
crossover temperature $\Delta=t\exp(-\pi v/g)$ (or the spin gap at the Fermi
level)\cite{flavor-inter}. 
In this Letter, we decouple 
${\cal H}_I={\cal H}_{bs}+{\cal H}_{holon}$, with:
\begin{eqnarray}
\label{BS}
{\cal H}_{bs}&\propto& -g\hbox{\large(}\cos{\sqrt{8}}\phi_{os}(x)+
\cos{\sqrt{8}}\phi_{\pi s}(x)\hbox{\large)},\\ \nonumber
{\cal H}_{holon}&\propto& -g\cos{\sqrt{8}}\Theta_{cf}(x).
\end{eqnarray}
This decoupling
does not affect the ground state properties and the gap equation.  
Moreover, the effective spin terms in Eq.(\ref{BS}) produce explicitly a 
{\it two-band Luther-Emery model}\cite{LE}. 
The model we propose here to describe is an impurity spin situated at the
edge $x=0$ of such d-wave superconductor. The bulk Hamiltonian obeys: 
$H=\int_0^L dx\ {\cal H}$. In the thermodynamic limit
we have a semi-infinite system.

First, it is convenient to rewrite the spin
part $H_{os}+H_{bs}$ on the entire axis $-\infty<x<\infty$, 
redefining $\Psi_{Rj\alpha}(x)=\hat{{\Psi}}_{j\alpha}(x)$ and
$\Psi_{Lj\alpha}(x)=\hat{{\Psi}}_{j\alpha}(-x)$, with
$\hat{{\Psi}}_{j\alpha}(x)\sim \exp i\Phi_{j\alpha}(x)$.
In terms of bosonic operators, one can simply equate (see Eq.(\ref{field})):
\begin{equation}
(\phi_{js}(x),\Theta_{js}(x))=
\frac{1}{2}\hbox{\Large(}\Phi_{js}(x)\mp\Phi_{js}(-x)\hbox{\Large)}.
\end{equation}
Rescaling the variables in a standard manner
$\phi_{js}\rightarrow\sqrt{K}\phi_{js}$
and $\Theta_{js}\rightarrow
\Theta_{js}/\sqrt{K}$, the (bulk) spin Hamiltonian becomes:
\begin{eqnarray}
H_{os}&=&\frac{v}{4}\int_{-\infty}^{+\infty} dx\
(\partial_x\Phi_{os})^2+(\partial_x\Phi_{\pi s})^2\\ \nonumber
H_{bs}&=&-\frac{g}{(2\pi a)^2}\int_{-\infty}^{+\infty} dx\ 
e^{i\sqrt{2K}\Phi_{os}(x)}
e^{-i\sqrt{2K}\Phi_{os}(-x)}e^{-i\pi K sgn(x)}+(o\rightarrow \pi).
\end{eqnarray}
For a complete introduction on bosonization with boundaries, consult for
instance ref.\cite{FG}. Second, in the band basis the Kondo interaction reads
\begin{equation}
H_k=\frac{J}{2}\sum_{j;\alpha\beta} \hbox{lim}_{x\rightarrow 0}
\{\hat{\Psi}_{j\alpha}^{\dag}(x)\sigma_{\alpha}^{\beta}
\hat{\Psi}_{j\beta}(x)\}\cdot\vec{S}=
J\sum_{j;\alpha\beta} \hbox{lim}_{x\rightarrow 0^+}
\{\Psi_{Rj\alpha}^{\dag}(x)\sigma_{\alpha}^{\beta}
\Psi_{Rj\beta}(x)\}\cdot\vec{S}.
\end{equation}
Backward scattering processes are already implemented by the open boundary:
$[\Psi_{Lj\alpha}(0)+\Psi_{Rj\alpha}(0)]=0$\cite{Gogolin}.
Using preceding transformations, the electron fields of Eq.(\ref{field})
read:
\begin{equation}
\Psi_{Pj\alpha}(x)=\hat{\Psi}_{j\alpha}(Px)=\kappa_{j\alpha}e^{i\hbox{\Large(}
P\frac{\sqrt{K}}{2}(\Phi_{j\alpha}(x)-\Phi_{j\alpha}(-x))+\frac{1}{2\sqrt{K}}
(\Phi_{j\alpha}(x)+\Phi_{j\alpha}(-x))\hbox{\Large)}}.
\end{equation}
The vanishing of the wave function at the origin gives also:
\begin{eqnarray}
\hbox{lim}_{x\rightarrow
  0^+}\{\Phi_{j\alpha}(x)-\Phi_{j\alpha}(-x)\}&=&\pi/\sqrt{K}\\ \nonumber
\hbox{lim}_{x\rightarrow
  0^+}\{\Phi_{j\alpha}(x)+\Phi_{j\alpha}(-x)\}&=&
2\Phi_{j\alpha}(0)-\pi/\sqrt{K}, 
\end{eqnarray}
and\cite{FG}:
\begin{equation}
\label{fieldn}
\Psi_{Rj\alpha}(0)=\hbox{lim}_{x\rightarrow
  0^+}\hat{\Psi}_{j\alpha}(x)
\simeq\kappa_{j\alpha}e^{i\frac{\Phi_{j\alpha}(0)}{\sqrt{K}}}\cdot
\end{equation}
Therefore, the two-channel Kondo interaction takes the form
\cite{remark}:
\begin{equation}
H_k=-\frac{J_z}{2\pi\sqrt{K}}S^z\partial_x\Phi_s(0)+\frac{J}{\pi
  a}\cos\Phi_{sf}(0)\exp(i\Phi_s(0))S^{+}+{\cal O}(\frac{gJ}{\pi v})+h.c.
\end{equation}
We have canonically transformed
$\Phi_s=(\Phi_{os}+\Phi_{\pi s})/\sqrt{2}$ and
$\Phi_{sf}=(\Phi_{os}-\Phi_{\pi s})/\sqrt{2}$.
The charge degrees of freedom are not coupled to the impurity, so we 
subsequently omit them in the following.
Moreover, it is convenient to use the unitary
rotation: ${\cal U}=\exp(i\Phi_s(0)S^z)$. We can first absorb the z-component
at the particular point $J_z=\pi v\sqrt{K}$, namely the generalized
Emery-Kivelson point. Then, the planar
Kondo term is transformed as in the usual
two-channel Kondo model\cite{EK,Giam}:
\begin{equation}
\tilde{H}_k={\cal U} H_k {\cal U}^{\dag}=\frac{J}{\pi a}\cos\Phi_{sf}(0)\cdot 
S^x+h.c.
\end{equation}
Leading order in $g$, the backscattering Hamiltonian becomes:
\begin{equation}
\tilde{H}_{bs}=H_{bs}\propto
+g\int_{-\infty}^{+\infty} dx\
e^{i\Phi_s(x)}e^{-i\Phi_s(-x)}[e^{i\Phi_{sf}(x)}e^{-i\Phi_{sf}(-x)}+
e^{-i\Phi_{sf}(x)}e^{i\Phi_{sf}(-x)}].
\end{equation}

Finally, we must retransform the bosonic variables as
spinless fermions. Using the Fermi
operators $\Psi_s\sim\exp i\Phi_s(x)$ and $\Psi_{sf}\sim\exp 
i\Phi_{sf}(x)$\cite{EK,Giam}, one arrives at
\begin{eqnarray}
\tilde{H}_{bs}&=&-g\int_{-\infty}^{+\infty} dx\  
\Psi_s^{\dag}(x)\Psi_s(-x)\Psi_{sf}^{\dag}(-x)\Psi_{sf}(x),\\ \nonumber
&=&-g\int_{-\infty}^{+\infty}
dx\ ic\ \Psi_s^{\dag}(x)\Psi_s(-x)sgn(x)+(s\rightarrow sf).
\end{eqnarray}
Using ref.\cite{c-exp}, one definitely obtains
$\tilde{H}=H_{os}+\tilde{H}_{bs}+\tilde{H}_k$, with
\begin{equation}
H_{os}+\tilde{H}_{bs}=
\sum_{j=s,sf}\int_{-\infty}^{+\infty} dx\
[v\Psi^{\dag}_j(x)(-i\partial_x)\Psi_{j}(x)-i\Delta\Psi^{\dag}_j(x)\Psi_j(-x)sgn
(x)],
\end{equation}
\begin{equation}
\tilde{H}_k=\frac{J}{\sqrt{\pi a}}[\Psi_{sf}(0)+\Psi_{sf}^{\dag}(0)]\hat{a}+
h.c.
\end{equation} 
$S^x=\hat{a}/\sqrt{2}$, $S^y=\hat{b}/\sqrt{2}$ 
and $S^z=i\hat{a}\hat{b}$. The $\hat{a}$ and $\hat{b}$ are Majorana (real) 
fermions.
Thus, we arrive at a resonant-level
type model that is solvable because it combines the Emery-Kivelson solution 
of the two-channel Kondo model with the
two band Luther-Emery model. Now
 we introduce another important energy scale, the bare Kondo
resonance\cite{EK,Giam,re}: $\Gamma=J^2/\pi v a$. 
It is
instructive to diagonalize first the bulk Hamiltonian, similarly
as in ref.\cite{Gogolin}. Bulk solutions for quasiparticles are plane
waves $\exp\pm ipx$, where $\epsilon_p=\pm[(vp)^2+{\Delta}^2]^{1/2}$.
But, starting with a semi-infinite system should produce novel Kondo features.
Precisely, we have new solutions at the edge $x=0$ 
(`bound S=1/2
spinons') with $\epsilon_p=0$ i.e. $p=\pm i\Delta/v$. Conversely
to usual superconductors
these zero-energy bound states exist without any proximity effect
with a normal metal\cite{Fauchere}. The
impurity Green's functions take the forms:
\begin{eqnarray}
A(\tau)&=&-\langle T{
  \hat{a}(\tau)\hat{a}(0)}\rangle=\frac{1}{\beta}\sum_{w_n}\frac{e^{i
  w_n\tau}}{iw_n-\Lambda(iw_n)}\\ \nonumber
B(\tau)&=&-\langle T{
  \hat{b}(\tau)\hat{b}(0)}\rangle=\frac{1}{\beta}\sum_{w_n}\frac{e^{i
  w_n\tau}}{iw_n}=\frac{1}{2}sgn(\tau),
\end{eqnarray}
$w_n$ are fermionic Matsubara frequencies and
$\Lambda(iw_n)=\Gamma(\sqrt{{w_n}^2+{\Delta}^2}+\Delta)/iw_n$.

{\it Physics
and conclusions}.--- 
A new interesting physics arises when $\Gamma\ll 2\Delta$. The
normal quasiparticles definitely disappear from the 
spectrum at low energy\cite{dlike}. A zero-energy spinon, from the $sf$-sector,
should form a stable bound state with {\it half} of the
impurity spin: the associated (rescaled) Kondo energy scale is
$w_{coh}=(2\Gamma\Delta-{\Gamma}^2)^{1/2}$, solution 
of $iw_n=\Lambda(iw_n)$. 
The situation becomes completely
different from that in a two-dimensional d-wave superconductor where
quasiparticles near the nodes mainly screen the 
impurity. In that case, both $\chi_{imp}$ and $C_{imp}$ 
go to zero at $T=0$ meaning that the impurity is 
completely screened.
Furthermore, the
reason behind the existence of $J_c$ is that the effective interaction between
the impurity and the normal excitations is momentum dependent and that it
vanishes at small momenta\cite{Cassanello_d-wave_Kondo}. In our case, an
overscreened Kondo effect can already take place when 
$J\rightarrow 0$.
When $T\ll\Gamma$, integrating over the coupling constant
the free energy takes the form:
\begin{equation}
\label{free}
F=F_o+F(\hat{a})=F_o-\frac{T}{2}\ln\cosh\frac{w_{coh}}{2T}\cdot
\end{equation}
For $J=0$, we have $S_o=-\partial F_o/\partial T=\ln 2$ because an isolated
impurity has a two fold degenerate ground state. Then, expanding the second
term of Eq.(\ref{free}) near $T\rightarrow 0$, the result is $S(\hat{a})=-\ln 
2/2$.
We find the same residual entropy $S(0)=S_o+S(\hat{a})=\ln 2/2$
as in the usual two-channel Kondo model (in a
metal)\cite{EK,Giam}: only half of the impurity is screened at zero
temperature. Furthermore, the
impurity contribution to the specific heat comes from thermal
activation, i.e.
$C_{imp}\propto (1/T^2)\cdot\exp-w_{coh}/T$\cite{Gogolin}. 
Another important point is that
performing a deviation from the generalized
Emery-Kivelson limit does not change much this
conclusion: the Green's function for a bound spinon is of the form
of $B(\tau)$, and then no $\ln$-divergence may arise in the second order
correction (in $\lambda=J_z-\pi v\sqrt{K}$) to the free energy.
We can also compute the local impurity 
susceptibility (i.e. when the magnetic field acts only on the
impurity spin)\cite{Go_Ners_Tsve}.
When $T\ll\Gamma$, it is convenient to use the following
asymptotic form of the propagator $(0\ll\tau\ll\beta)$,
\begin{equation}
A(\tau)=\frac{1}{2}e^{-w_{coh}\tau}.
\end{equation}
This leads to,
\begin{equation}
\label{qui}
\chi_{l-imp}(T)=\int_{{\tau}_o}^{\beta} d\tau\ \langle
S^z(\tau)S^z(0)\rangle 
\propto\frac{1}{w_{coh}}\{e^{-w_{coh}\tau_o}-e^{-w_{coh}/T}\}\cdot
\end{equation}
 The exponential contained in $\hat{a}$ correlations cuts the $\tau$-integral
off $(\tau_o\sim 1/\Gamma)$. The most interesting observation
is the absence of a $\ln$-divergence in $\chi_{l-imp}$ at zero temperature,
despite the splitting of the impurity into two Majorana fermions
$\hat{a}$ and $\hat{b}$\cite{Go_Ners_Tsve}. It is noteworthy
that {\it the
splitting
phenomenon produces only a finite maximum in the local impurity susceptibility
when $T\rightarrow 0$}.
When $\Gamma\sim 2\Delta$, the bound state formed
between the Majorana fermion $\hat{a}$ and an 
edge spinon disappears 
i.e. $w_{coh}=0$ and for
larger couplings only normal quasiparticles will be involved in the
Kondo coupling with the impurity. At low temperatures, this results in
$C_{imp}(T)\rightarrow 0$ and $\chi_{l-imp}(T)\rightarrow 1/T$
(obtained by expanding Eq.(\ref{qui}) near $w_{coh}=0$
with $\tau_o\rightarrow 0$). 
We have a crossover region where
the impurity is completely unscreened; the density of states
at the Fermi level tends to zero and
quasiparticles cannot pretend to form
a stable bound state with the impurity. Finally, in the
limit $\Delta\rightarrow 0$ (which corresponds to K=1), the density of states
at the zero-energy level becomes 
{\it finite} and we recover previous results:
$\Lambda(iw_n)=-i\Gamma sgn(w_n)$ and then 
$\chi_{l-imp}(T)\propto\ln T$,
$C_{imp}(T)\propto T\ln T$\cite{Giam}. The impurity is now overscreened by
two channels of normal electrons.

To conclude, a solvable model describing a magnetic impurity coupled to bound
spinons located at the edge of a one dimensional
 d-wave superconductor, is proposed in this Letter. We find a new type
of two-channel Kondo effect resulting in a quite unusual low-temperature
thermodynamics.

I thank the theoretical group of ETH for stimulating discussions, and
especially U. Ledermann and M. Rice for collaboration on a related
work. This paper is devoted to my father.


\begin{thebibliography}{10}
\bibitem{QP}
The onset of Kondo screening at a critical value of the exchange coupling
constant is a quantum critical point.
\bibitem{newr}
At low temperatures, spin Cooper scatterings {\it within a band} are also 
flowing (marginally) to strong
couplings. On the other hand, 
we can omit them because they give $\phi_{js}\sim 0$,
that is also produced by ${\cal H}_I$.
\bibitem{flavor-inter}
Here, we have {\it implicitly} interchanged $K\leftrightarrow K^{-1}$
in the definition of the Luttinger parameter in the flavor sector $cf$ 
(to be totally consistent with ref.[10]).
\bibitem{dlike}
One should not overinterpret the `d-wave' symmetry. In the two-band
model, it is just equivalent to: $\Delta_o\Delta_{\pi}^{\dag}<0$. Nodes
in such system are not clearly defined.
\bibitem{remark}
Since $g\ll v$, terms in ${\cal O}(gJ/\pi v)$ give very small corrections 
to the Kondo
resonance and will be subsequently neglected.
\bibitem{c-exp}
$c=-i\langle\Psi^{\dag}_s(-x)\Psi_s(x)\rangle=+i\langle
\Psi^{\dag}_s(x)\Psi_s(-x)\rangle$. It obeys the self-consistent
equation: $c=(gc/\pi v)\ln(t/gc)$ and then, $(gc)=\Delta$, the superconducting
energy scale.
\bibitem{re}
Since $J_z\gg J$, $\Gamma$ should not be confused with the Kondo
scale of the isotropic model.
\end{thebibliography}
\end{document}